# Radiostrontium activity concentrations in milk in the Republic of Croatia for 1961 - 2001 and dose assessment

Z. Franic[#], N. Lokobauer and G. Marovic[*]

ABSTRACT

Results of systematic measurements of $^{90}$Sr activity concentrations in milk for the period 1961 - 2001 are summarized. An exponential decline of radioactivity followed the moratorium on atmospheric nuclear testing. The highest activity of $^{90}$Sr deposited by fallout, being 1060 Bq m$^{-2}$, was recorded in 1963, while the peak $^{90}$Sr activity concentration in milk, 1.42 ± 0.17 Bq L$^{-1}$, was recorded in 1964. The values in year 2001 for fallout deposition and milk were 7.7 Bq m$^{-2}$ and 0.07 ± 0.03 Bq L$^{-1}$, respectively. The reactor accident at Chernobyl caused higher $^{90}$Sr levels only in 1986. $^{90}$Sr fallout activity affects milk activity, the coefficient of correlation between $^{90}$Sr fallout activity and $^{90}$Sr activity concentrations in milk being 0.80. The transfer coefficient from fallout deposition to milk was estimated to be $2.5 \times 10^{-3}$ Bq y L$^{-1}$ per Bq m$^{-2}$. The dose incurred by milk consumption was estimated for the Croatian population, the annual collective effective dose in 2001 being approximately 2.0 man-Sv.

Key words: $^{90}$Sr; milk; contamination; environmental; fallout; dose

[#]From 2002 to 2004 at the position of Vice-Minister of Science and Technology in the Republic of Croatia, Trg J. J. Strossmayera 4, HR-10000 Zagreb, Croatia.

[*]Institute for Medical Research and Occupational Health, Ksaverska cesta 2, PO Box 291, HR-10001 Zagreb, Croatia.

For correspondence or reprints contact: Z. Franic at the address of the Institute, or email at franic@franic.info

INTRODUCTION

The dominant route for the introduction of artificial radionuclides into the environment until the nuclear accident in Chernobyl, Ukraine, on 26 April 1986, has been the radioactive fallout resulting from atmospheric nuclear weapon tests. Atmospheric nuclear explosions have been conducted since 1945 and were specially intensive in the 1960s, i.e., before a moratorium on atmospheric nuclear tests became effective. However, similar, but smaller tests were conducted by the Chinese and French in the 1970s until 1980. The radioactivity of most environmental samples, including those that enter human food chain, is usually in good correlation with fallout activity, i.e., surface deposit in Bq m$^{-2}$ (UNSCEAR 1982; UNSCEAR 1988). Among the anthropogenic radionuclides present in global fallout, $^{137}$Cs and $^{90}$Sr have been regarded as the fission products of a major potential hazard to living beings due to the unique combination of their relatively long half-lifes, and their chemical and metabolic properties resembling those of the potassium and calcium, respectively. Milk, containing both potassium and calcium is, therefore, the sensitive indicator for presence of fission products in the environment. In addition, milk as the very important foodstuff in Croatian dietary habits, is potentially a major source of radioactive contamination. Consequently, investigations of radiostrontium and radiocaesium in dairy milk take significant part of an extended and still on going monitoring program of radioactive contamination of the environment in Croatia (Popovic 1963 - 1978; Bauman et al. 1979 - 1992; Kovac et al. 1993 - 1998; Marovic et al. 1999 - 2002.). In this paper we have summarized the results of long-term systematic measurements of radiostrontium activity in milk.

MATERIAL AND METHODS

Fallout samples were collected monthly in the city of Zagreb at the location of the Institute for Medical Research and Occupational Health (45°50' N, 15°59' E). The other fallout sampling locations are the cities of Zadar on the Adriatic coast (44°06' N, 15°15' E) and Osijek in eastern Croatia (45°33' N, 18°42' E). The funnels used for fallout collection had 1 m$^2$ area. The precipitation amount was measured by Hellman pluviometer. In order to account for dry deposition (i.e., activity deposited by particle settling and other "dry" processes) for periods when there was no precipitation, funnels were rinsed by 1 L of distilled water (U.S. Department of Health, Education and Welfare 1967, U.S. Department of Energy, 1957 - 1997.)

Samples of milk, produced by Zagreb dairy, were purchased on daily basis in a quantity of 1 L. The Zagreb diary collects milk from local farmers all over Croatia, and after processing supplies milk for majority of Croatian population through markets and supermarkets. Cumulative monthly samples were analyzed monthly by taking aliquot. In the 1960s, $^{90}$Sr was determined by the conventional radiochemical analysis with fuming nitric acid separation. From 1970 to the present, $^{90}$Sr was determined by extraction with tributyl phosphate, except in the year of the Chernobyl nuclear accident, when fuming nitric acid was also used.

After the radiochemical treatment of fallout and milk samples (U.S. Department of Energy, 1957 - 1997), the radioactivity of $^{90}$Sr was determined by counting the beta emission from its decay product, $^{90}$Y, in a low-background, anti-coincidence, shielded Geiger-Müller counter. Counting time depended on $^{90}$Sr activity concentration in samples, but was never less than 60,000 s, typically being 80,000 s. The values of ± 2σ counting errors were approximately 10% of the reported values of activity concentrations. The efficiency calibration was carried out using sources provided by the International Atomic Energy Agency (IAEA). Quality control is performed on monthly basis by measuring blank samples and standards. Quality assurance and intercalibrations of radioactivity measurements were performed through participation in the IAEA and WHO international quality control programs.

RESULTS AND DISCUSSION

*Fallout*

The highest recorded $^{90}$Sr activity deposited in Croatia, 1060 Bq m$^{-2}$, was recorded in Zagreb in 1963. Ever since that time $^{90}$Sr fallout activities decreased exponentially, until the nuclear accident in Chernobyl that caused a minor increase of $^{90}$Sr activity in fallout. However, due to the prevailing meteorological conditions at the time after the accident that influenced the formation and direction of Chernobyl plumes, in Croatia the $^{90}$Sr peak (surface deposition being about 210 Bq m$^{-2}$) was recorded only in fallout collected in the city of Zagreb, which was affected by the edge of the plume as indicated on the Figure 1. In fallout samples collected on other places in Croatia (Zadar and Osijek), $^{90}$Sr activity concentrations were not detected.

**Figure 1 about here**

Unlike the debris from the atmospheric testing of nuclear weapons, the radionuclides that originated from the Chernobyl accident were not released directly into the upper atmosphere. As the result of the release mechanism and the meteorological conditions, the refractory components of the Chernobyl debris (e.g., $^{90}$Sr) were deposited closer to the accident location than the more volatile constituents (e.g., radiocaesium) (UNSCEAR 1988; Aarkrog 1988). Thus, $^{90}$Sr was not subjected to the global dispersion processes, and was deposited on the Earth's surface within a period of a few days to a few weeks after the accident. In addition, the late spring and early summer of 1986 in Croatia were rather dry, which was especially true for the Adriatic region, leading to relatively low direct radioactive contamination of environment (Franic and Bauman 1993; Franic et al. 1999). Consequently, due to much lesser volatility of strontium compared to caesium, the nuclear accident at Chernobyl did not cause a major increase in $^{90}$Sr activity in environmental samples in Croatia, contrary to radioactive isotopes of caesium. For example, in May 1986 in the Zagreb area $^{137}$Cs surface deposit was 6200 Bq m$^{-2}$.

The increased $^{90}$Sr fallout activity concentrations were recorded mainly in May 1986, leading to the surface deposit of 211.3 Bq m$^{-2}$ in Zagreb area (Franic 1994). In 1987, the $^{90}$Sr fallout activity concentrations dropped to 12.7 Bq m$^{-2}$, i.e., the pre-Chernobyl value, while $^{137}$Cs activity concentrations remained elevated. The $^{137}$Cs:$^{90}$Sr activity ratio that for the pre-Chernobyl period in most of the environmental samples collected in Croatia was relatively constant, generally ranging between values 1 and 3, after the Chernobyl accident was notably altered (Popovic 1963-1978; Bauman et. al. 1979-1992). In 1986, this ratio for milk had a value of 107 and similar ratio values were determined in most of the other environmental samples in Croatia.

*Milk*

The data for the measurements of radiostrontium activities in milk and fallout (1961 - 2001) are presented on figures 2 and 3. The peak value of $^{90}$Sr in milk, 1.42 ± 0.17 Bq L$^{-1}$, was recorded in 1964, after the most intensive nuclear weapon tests. As in the case of fallout, $^{90}$Sr levels in milk were exponentially decreasing from 1964 to 1986, when the Chernobyl accident caused a $^{90}$Sr activity concentration peak of 0.43 ± 0.41 Bq L$^{-1}$. The high standard deviation in year 1986 was caused by very high activity concentrations in May and June, being 1.52 and 1.00 Bq L$^{-1}$ respectively. After the Chernobyl accident, the $^{90}$Sr activity concentrations in milk very quickly decreased to pre-Chernobyl values, the activity concentration in 1987 being $0.17 \pm 0.18$ Bq L$^{-1}$. For comparison, in 1985, the recorded activity concentration was $0.16 \pm 0.02$ Bq L$^{-1}$. For the overall observed period, the lowest $^{90}$Sr milk activity concentration was recorded in 2000, being $0.06 \pm 0.02$ Bq L$^{-1}$, while in 2001 it was $0.07 \pm 0.03$ Bq L$^{-1}$.

By regression analysis it can be demonstrated that the $^{90}$Sr fallout activity concentrations are related to milk activity concentrations, the coefficient of correlation being 0.80. Therefore, from fallout data $^{90}$Sr activity concentrations in milk can be very simply modeled as:

$$A_{milk}(t) = 0.00127 \times A_{fallout}(t) + 0.219 \qquad (1)$$

where:

$A_{milk}(t)$ is the $^{90}$Sr activity concentration in milk in Bq L$^{-1}$ and

$A_{fallout}(t)$ is the $^{90}$Sr fallout activity concentration in Bq m$^{-2}$.

The data on $^{90}$Sr in milk, fallout and model (1) are shown on Figure 2. This simple regression systematically underestimates observed milk data in 1960s and 1970s, and overestimates afterwards.

**Figure 2. about here**

To assess the $^{90}$Sr transfer from fallout to milk the mathematical model for food products recommended by UNSCEAR (UNSCEAR 1982) was applied. This model was previously tested on milk samples collected in Croatia (Lokobauer 1984). The function has the following form:

$$A_k(t) = b_1 U_k(i) + b_2 U_k(i-1) + b_3 \sum_{m=1}^{\infty} e^{-\mu m} U_k(i-m) \qquad (2)$$

where:

| | |
|---|---|
| $A_k(t)$ | is the activity concentration of radionuclide $k$ in food (milk), the unit for $A_k(t)$ being Bq L$^{-1}$, |
| $U_k(i)$ | is the fallout deposition rate of radionuclide $k$ in a year $i$ (Bq m$^{-2}$ y$^{-1}$), |
| $\Sigma e^{-\mu m}...$ | is the cumulative fallout deposit for radionuclide $k$ as the result of deposition in previous years (Bq m$^{-2}$ y$^{-1}$), |
| $\mu^{-1}$ | is the effective (i.e., observed) mean residence time of available $^{90}$Sr in soil, |
| $b_1, b_2, b_3$ | are the factors which can be derived from reported data by regression analysis. The unit is Bq L$^{-1}$ / (Bq m$^{-2}$ y$^{-1}$). |

The equation (2) assumes the chain model for the transfer of radionuclides between environmental compartments ($C_0 ... C_5$), linking the input to the atmosphere to the dose in man:

$$\text{Input } (C_0) \rightarrow \text{Atmosphere } (C_1) \rightarrow \text{Earth's surface } (C_2) \rightarrow \text{Diet } (C_3) \rightarrow \text{Tissue } (C_4) \rightarrow \text{Dose } (C_5)$$

The physical meaning of terms in model (2) is as follows: the first term (*rate factor*) describes direct deposition and transfer, the second term (*lag factor*) describes contamination through fallout from previous year, and the third term (*soil factor*) reflects the contamination from fallout deposition accumulated from all preceding years, the exponential describing the combined physical decay and any other decrease in availability of considered radionuclide due to various other processes (like penetration in deeper soil layers, dilution etc.).

Regression analysis gives for the factors $b_1$, $b_2$ and $b_3$ values $1.00 \times 10^{-4}$, $9.00 \times 10^{-5}$ and $1.23 \times 10^{-3}$ Bq L$^{-1}$ / (Bq m$^{-2}$ y$^{-1}$) respectively, and for the constant $\mu$ the value of 0.43 y$^{-1}$, the coefficient of correlation between actual data and data predicted by model (2) being 0.77. Compared with the simple regression model given by equation (1) it is just slightly lesser. However, model (2) has physical meaning, contrary to model (1) which just demonstrates that the milk activity is the temporal function of fallout activity. The reciprocal value of the constant $\mu$, 2.3 y, is the effective mean residence time of available $^{90}$Sr in soil. However, to find the real residence time, $\mu_R$, constant $\mu$ should be corrected for the radioactive decay. Therefore:

$$\mu = m + \mu_R \qquad (3)$$

where $\ln(2)/m = 29.1$ y is the half-life of $^{90}$Sr.

From equation (3), the real mean residence time for $^{90}$Sr in soil ($T_M = 1/\mu_R$), was found to be 2.5 y, which is just slightly higher than effective mean residence time. Mean residence time of 2.5 y is much smaller from the value for the northern hemisphere found in literature. Namely, on the northern hemisphere, the average value for the factor $\mu$ is 0.12 y$^{-1}$ (WHO 1983), which means that the effective mean residence time and the real mean residence time of available $^{90}$Sr have respective values of 8.3 y and 10.4 y. However, it should be noted that residence time of available $^{90}$Sr in soil, varies for individual foods and soil conditions (WHO 1983), and in some cases can be several times smaller than the northern hemisphere average.

The effective mean residence time for $^{90}$Sr can be found by fitting the milk data to a simple exponential function:

$$A_{milk}(t) = A_{milk}(0)\, e^{-bt} \qquad (4)$$

The effective mean residence time based on the reciprocal value of constant b resulting from the fit, equals 10.7 years. By correcting for the radioactive decay using equation analog to (3), for the real $^{90}$Sr mean residence time in milk is obtained 14.6 years, which is nearly 6 times longer than for the mean residence time for $^{90}$Sr in soil calculated by model (2). It should be noted that the mean residence time in milk, estimated that way, is only an indicator of the rate by which the availability of $^{90}$Sr decreases with time and is not related to the metabolic behavior of strontium in cows. The $^{90}$Sr activity concentrations in milk and the fit obtained by equation (2) are shown in Figure 3.

**Figure 3. about here**

The transfer between successive steps in the pathway chains is described by transfer coefficients, which relate infinite time integrals of activity concentration in the relevant compartments. Thus, $P_{23}$, the transfer coefficient from fallout deposition (compartment 2) to diet (compartment 3) is given by the following equation:

$$P_{23} = \frac{\int_0^\infty A_k(t)\,dt}{\int_0^\infty U_k(t)\,dt} \qquad (5)$$

where:

$A_k(t)$ is the activity concentration of radionuclide $k$ in food, i.e., milk (BqL$^{-1}$) and

$U_k(t)$ is the fallout deposition rate of radionuclide (Bqm$^{-2}$y$^{-1}$).

Using values of $A_k(t)$ and $U_k(t)$ assessed on the yearly basis, the integration can be replaced by a summation. The combination of equations (2) and (5) leads to:

$$P_{23} = b_1 + b_2 + b_3 \frac{e^{-\mu}}{1 - e^{-\mu}} \qquad (6)$$

$P_{23}$ for $^{90}$Sr in milk was assessed to be $2.5 \times 10^{-3}$ Bq y L$^{-1}$/(Bq m$^{-2}$). That means that with each Becquerel deposited by fallout on an area of 1 m$^2$ of soil, the activity concentration of 1 L of milk increases by $2.5 \times 10^{-3}$ Bq. For comparison, the transfer coefficient $P_{23}$ for total diet was estimated to be $4 \times 10^{-3}$ Bq y kg$^{-1}$/(Bqm$^{-2}$) (UNSCEAR 1982).

*Doses incurred by milk consumption*

In the Republic of Croatia (4.4 millions inhabitants) consumption of milk, is approximately 87 L per year per every person (CBS 2001; CBS 2002). However, if all the dairy products are taken into account, than the effective consumption of milk significantly increases (CBS 2001; CBS 2002; Kolega 1994, Colic-Baric 2001, Colic-Baric et al. 2001; Colic-Baric and Brlecic 2001). Therefore, the consumption of milk and diary products can potentially lead to significant radiation doses. The effective dose (Sv) incurred due to milk consumption over a set period depends on the activity of radionuclides that are present in milk and on the quantity of milk consumed. The dose can be expressed as:

$$E = C \int \sum_k D_{ef}(k) A_k(t) dt \qquad (7)$$

where:

E    is the total effective dose in Sv,

C    total annual *per caput* consumption of milk,

$D_{cf}(k)$    the dose conversion factor for radionuclide $k$ and

$A_k$    the mean activity concentration of radionuclide $k$ in milk (Bq L$^{-1}$).

For equal time increments of one month, integral from equation (7) can be replaced by the respective sums of average monthly activity concentrations. $D_{cf}(k)$, the committed effective dose per unit intake (the dose conversion factor), relates the ingested activity concentration to the effective dose to the whole body. Calculated doses due to $^{90}$Sr in milk for the peak year (1964), pre-Chernobyl, Chernobyl, post Chernobyl year and year 2001 for various ages are presented in Table 1. For the dose conversion factors were taken ICRP values (IAEA 1996). The statistical data on milk consumption for different age groups were taken from the literature (Kolega 1994, Colic-Baric 2001, Colic-Baric et al. 2001; Colic-Baric and Brlecic 2001).

**Table 1. about here**

The total dose due to $^{90}$Sr in 2001 for adult was approximately 0.27 µSv. For comparison, in the same year the dose due to $^{137}$Cs was just slightly less 0.24 µSv, the effective dose per unit input for $^{137}$Cs being 1.3 × 10$^{-8}$ Sv Bq$^{-1}$ (IAEA 1996). The annual dose received by $^{90}$Sr intake via food in the mid 1990s was estimated to be about 3 µSv in Croatia (Lokobauer et al. 1998).

Using doses from table 1 and data on age distribution of Croatian population (CBS 2002), and assuming similar radiostrontium milk activity concentrations in other Croatian regions, the collective effective dose for Croatian population due to $^{90}$Sr ingestion by milk consumption in year 2001 can be estimated to be 2.0 man-Sv, while the annual collective dose in the year of Chernobyl was 13.0 man-Sv. The same year (1986), the collective effective dose for Croatian population incurred from milk due to $^{134}$Cs and $^{137}$Cs was estimated to be about 200 man-Sv, 59% of which was attributed to $^{137}$Cs and rest to $^{134}$Cs (Franic et al. 1998).

CONCLUSIONS

The $^{90}$Sr activity in milk decreases exponentially since 1960s, after the intensive atmospheric nuclear weapon test stopped. Peak values were recorded in 1964, following the most intensive nuclear weapon tests.

Doses due to radiostrontium from milk consumption are small, in spite of considerable consumption of milk by the Croatian population. The collective effective dose to Croatian population in 2001 from $^{90}$Sr ingestion of milk was 2.0 man-Sv.

It should be noted that although radiocaesium levels in the environment in the year of Chernobyl accident were much higher than those of radiostrontium in the following years the doses that Croatian population received from $^{137}$Cs and $^{90}$Sr were approximately the same. This can be explained by fact that $^{90}$Sr transfer from soil to the food chain is considerably more efficient, than for $^{137}$Cs.

Acknowledgment - This work received financial support from the Ministry of Science and Technology of the Republic of Croatia under grant # 00220204 (Environmental Radioactivity)

**Tables**

Table 1. Annual doses (μ Sv) received by $^{90}$Sr intake by consumption of milk calculated for various age groups.

| Age (years) | Average yearly milk consumption (L) | Effective dose conversion factor (Sv Bq$^{-1}$) | 1964 | 1985 | 1986 | 1987 | 2001 |
|---|---|---|---|---|---|---|---|
| | | | Observed milk activity concentrations (Bq L$^{-1}$) | | | | |
| | | | $1.42\times10^3$ | $1.59\times10^2$ | $4.31\times10^2$ | $1.66\times10^2$ | $6.58\times10^1$ |
| | | | Doses (μ Sv) | | | | |
| ≤ 1 | 21 | $2.3 \times 10^{-7}$ | $6.9 \times10^0$ | $7.7 \times10^{-1}$ | $2.1 \times10^0$ | $8.0 \times10^{-1}$ | $3.2 \times10^{-1}$ |
| 1 - 2 | 203 | $7.3 \times 10^{-8}$ | $2.1 \times10^1$ | $2.4 \times10^0$ | $6.4 \times10^0$ | $2.5 \times10^0$ | $9.7 \times10^{-1}$ |
| 2 - 7 | 251 | $4.7 \times 10^{-8}$ | $1.7 \times10^1$ | $1.9 \times10^0$ | $5.1 \times10^0$ | $2.0 \times10^0$ | $7.8 \times10^{-1}$ |
| 7 - 12 | 288 | $6.0 \times 10^{-8}$ | $2.5 \times10^1$ | $2.8 \times10^0$ | $7.4 \times10^0$ | $2.9 \times10^0$ | $1.1 \times10^0$ |
| 12 - 17 | 265 | $8.0 \times 10^{-8}$ | $3.0 \times10^1$ | $3.4 \times10^0$ | $9.1 \times10^0$ | $3.5 \times10^0$ | $1.4 \times10^0$ |
| adult | 149 | $2.8 \times 10^{-8}$ | $5.9 \times10^0$ | $6.7 \times10^{-1}$ | $1.8 \times10^0$ | $6.9 \times10^{-1}$ | $2.7 \times10^{-1}$ |

**Figures**

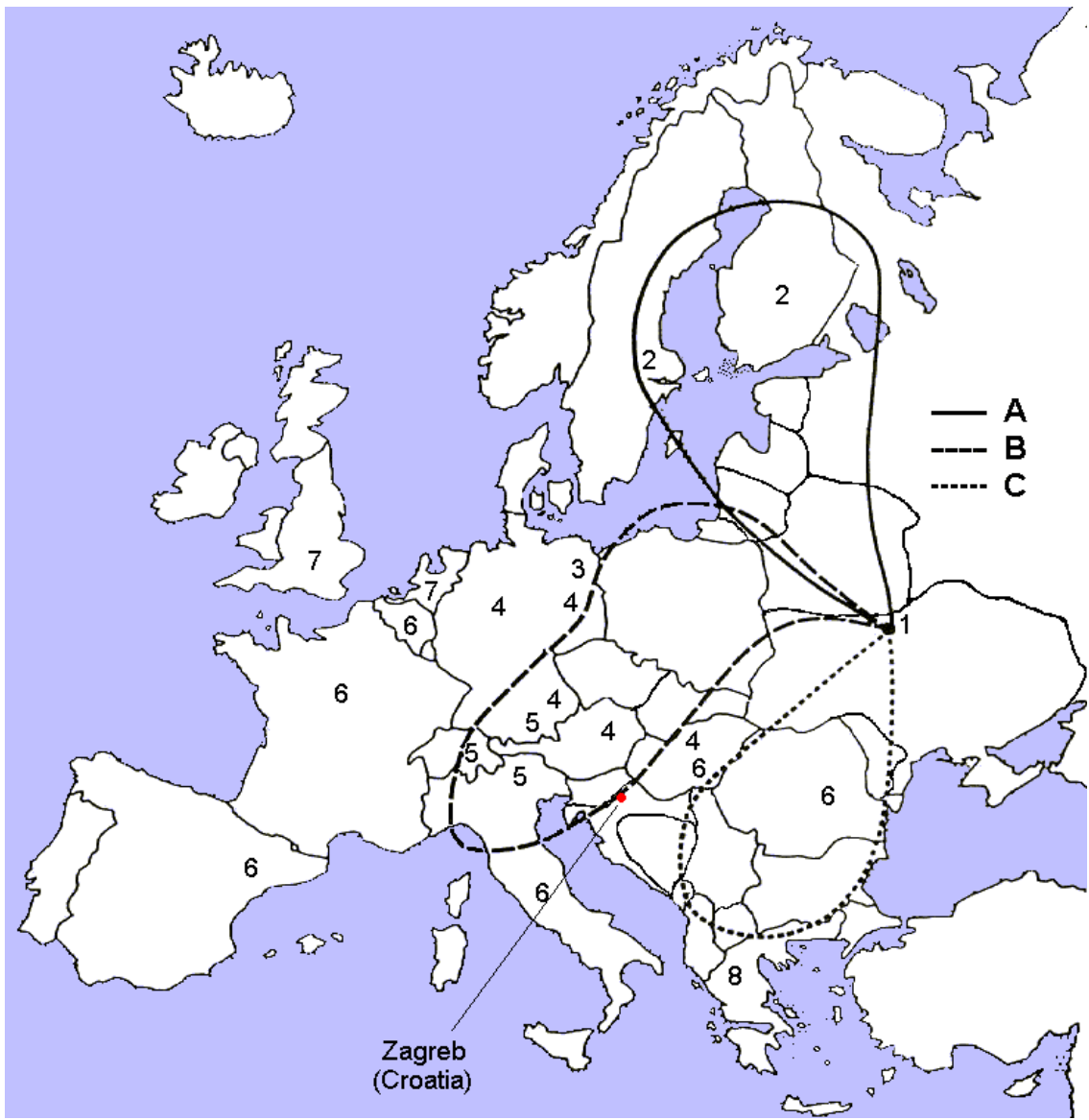

Figure 1    Spreading of radioactive plumes over the Europe after the Chernobyl nuclear accident. Numbers 1- 8 represent plume arrival times at respective areas: 1 = April 26, 2 = April 27, 3 = April 28, 4 = April 29, 5 = April 30, 6 = May 1, 7 = May 2 and 8 = May 3. The figure is taken from UNSCEAR report for 1988.

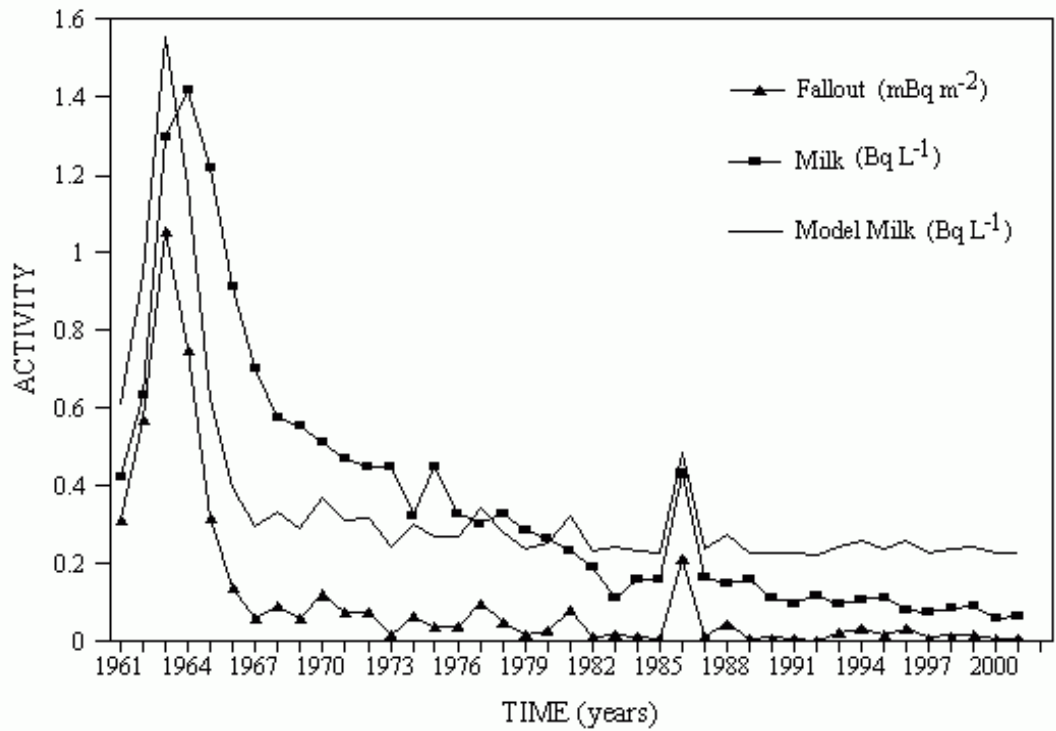

Figure 2    $^{90}$Sr fallout and milk activities and modeled milk activities using equation (1)

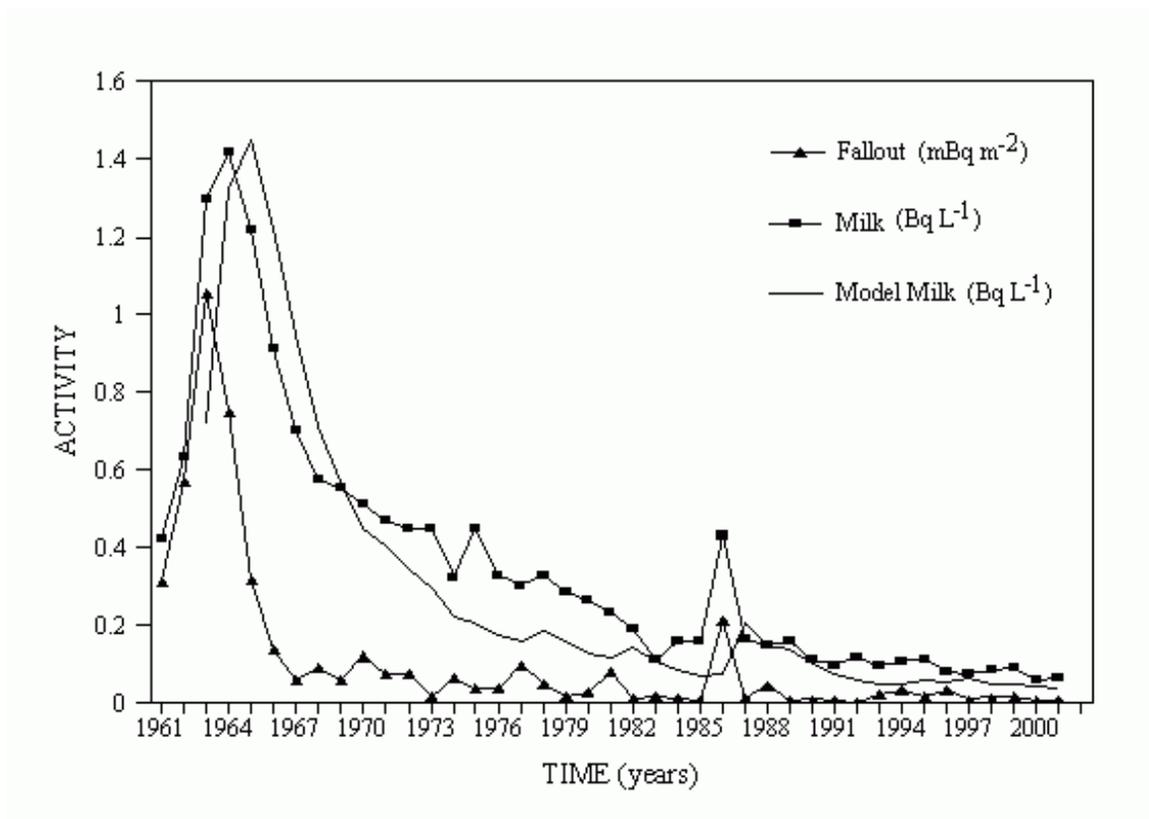

Figure 3    $^{90}$Sr fallout and milk activities and modeled milk activities using equation (2).